\pdfoutput=1
\documentclass[11pt]{article}

\usepackage[final]{acl}
\usepackage{times}
\usepackage{latexsym}

\usepackage[T1]{fontenc}
\usepackage[utf8]{inputenc}

\usepackage{microtype}

\usepackage{inconsolata}

\usepackage{graphicx}

\usepackage{booktabs}
\usepackage{multirow}
\usepackage{xcolor}
\usepackage{colortbl}
\usepackage{array}
\usepackage{amsmath}
\usepackage{amssymb}
\usepackage{tabularx}
\usepackage{makecell}
\usepackage{listings}
\usepackage{textcomp}
\usepackage{multicol}
\usepackage{float}
\usepackage{adjustbox}
\usepackage{enumitem}

\title{SATER: A Self-Aware and Token-Efficient Approach to Routing and Cascading}

\author{Yuanzhe Shen, Yide Liu, Zisu Huang, Ruicheng Yin\\
{\bf Xiaoqing Zheng\thanks{\ \ Corresponding author.}, Xuanjing Huang} \\
  School of Computer Science, Fudan University, Shanghai, China \\
  \texttt{\{yzshen25\}@m.fudan.edu.cn} \\
 \texttt{\{zhengxq,xjhuang\}@fudan.edu.cn} \\}

\begin{document}
\maketitle
\begin{abstract}
Large language models (LLMs) demonstrate remarkable performance across diverse tasks, yet their effectiveness frequently depends on costly commercial APIs or cloud services. Model selection thus entails a critical trade-off between performance and cost: high-performing LLMs typically incur substantial expenses, whereas budget-friendly small language models (SLMs) are constrained by limited capabilities. Current research primarily proposes two routing strategies: pre-generation routing and cascade routing. Both approaches have distinct characteristics, with cascade routing typically offering superior cost-effectiveness and accuracy despite its higher latency. To further address the limitations of both approaches, we introduce SATER, a dual-mode compatible approach that fine-tunes models through shortest-response preference optimization and a confidence-aware rejection mechanism. SATER significantly reduces redundant outputs and response times, while improving both the performance of pre-generation routing and the efficiency of cascade routing. Experiments across three SLMs and six datasets, varying in type and complexity, demonstrate that SATER achieves comparable performance while consistently reducing computational costs by over 50\% and cascade latency by over 80\%.\footnote{Our code is
available at \url{https://github.com/Lonely-tian/SATER}.} 

% Building on the intuitive assumption that simple tasks can be effectively managed by SLMs, with complex tasks allocated to high-performance LLMs, existing research primarily proposes two routing strategies: pre-generation routing and cascade routing.

\end{abstract}

\section{Introduction}

With the rapid advancement of large language models (LLMs) \citep{yang2024qwen2,liu2024deepseek}, their outstanding performance in diverse natural language processing tasks has solidified their role as a cornerstone of artificial intelligence applications. However, their operation entails significant computational costs, high energy consumption, and reliance on specialized hardware, creating substantial financial burdens and raising critical concerns about environmental sustainability \citep{kaack2022aligning, luccioni2024environmental} and technological accessibility. Consequently, optimizing efficiency and reducing resource consumption have become pivotal challenges \citep{varangot2025doing}.

Building on the intuitive assumption that simple tasks can be effectively managed by SLMs, with complex tasks allocated to high-performance LLMs, current research primarily proposes two approaches: pre-generation routing and post-generation routing (cascade routing). Pre-generation routing methods, such as HybridLLM \citep{dinghybrid} and RouteLLM \citep{ong2024routellm}, train classifiers to predict task complexity, avoiding generation overhead. In contrast, cascade routing evaluates response quality for decision-making, as seen in FrugalGPT \citep{chen2023frugalgpt}, which trains an answer correctness classifier, and AutoMix \citep{aggarwal2024automix} and MoT \citep{yuelarge}, which assess output confidence through multiple sampling. The latter demonstrates better adaptability across various tasks and generally outperforms pre-generation routing. Although cascade routing is more stable and superior, it requires a complete generation process. When SLMs' responses are rejected (common in complex tasks), regeneration is needed, and long responses introduce additional latency, leading to significant latency polarization and increased costs. These limitations suggest that the performance, cost-effectiveness, and latency of both routing approaches can be further optimized.

\begin{figure*}[t]
  \centering
  % \vspace{-.15in}
  \includegraphics[width=1.0\textwidth]{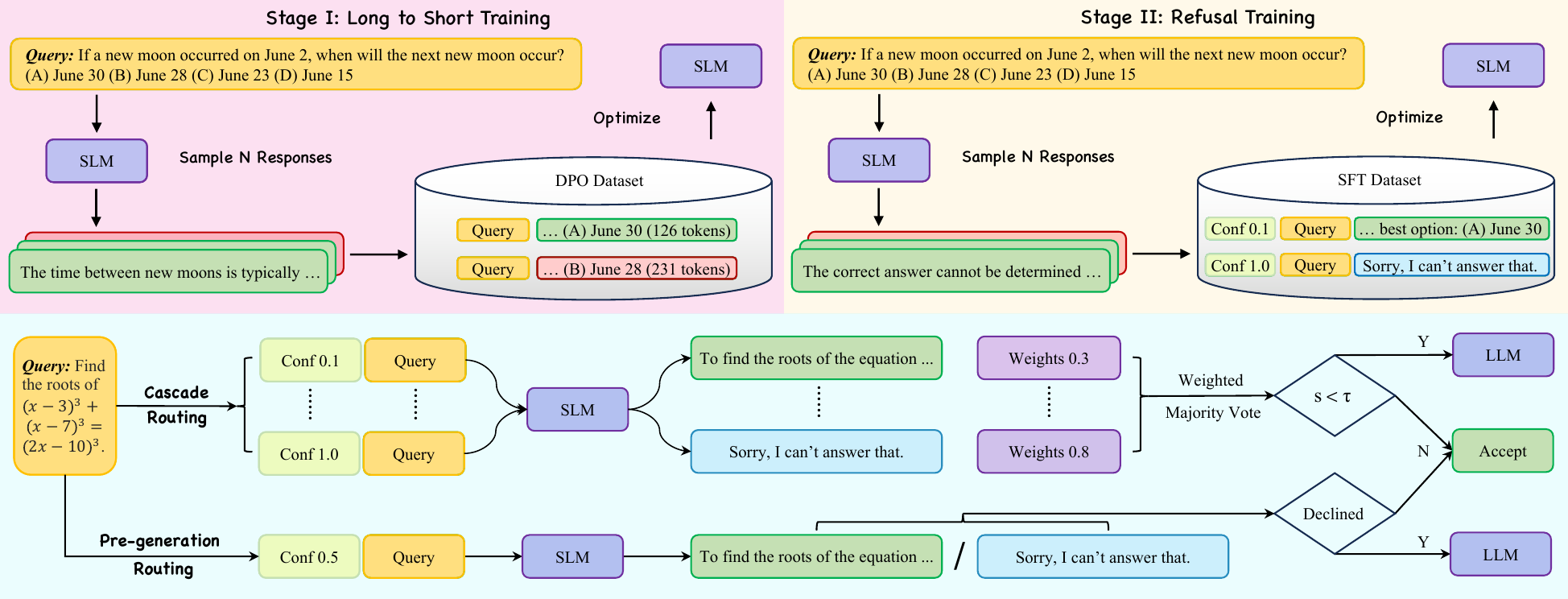}
  % \vspace{-.3in}
  \caption{Illustration of SATER. We train the SLM in two stages for optimal cost, accuracy, and latency. Stage I performs preference optimization with the shortest correct and longest incorrect responses, while Stage II employs prompt-based fine-tuning to teach the SLM to reject complex tasks. During inference, rejected queries are either routed directly to LLMs (pre-generation) or processed via weighted majority voting (cascade) for refined routing.}
  \label{score}
% \vspace{-.15in}
\end{figure*}

However, current evaluation frameworks for routing strategies present several limitations. In pre-generation routing, assessment results are highly susceptible to cases where both SLMs and LLMs fail to handle queries effectively. Furthermore, existing metrics such as cost-performance curves, APGR and CPT (as proposed in RouteLLM) are subject to ``Performance Gap Bias'' – a phenomenon where router performance appears inflated on benchmarks with minimal performance differentials between SLMs and LLMs. To overcome these limitations, we introduce two metrics: Tradeoff Area (ToA) and Tradeoff Gain Ratio (ToGR), designed to provide more robust evaluation of pre-generation routing strategies. For cascade routing, current methods typically rely on coarse estimates based on cost per million tokens and the number of samples. In the context of increasing emphasis on output length and generation efficiency, such simplifications fail to accurately capture the impact of generation length on cost and latency. To address this gap, we introduce an evaluation framework based on actual generation length, incorporating two new metrics: Average Generation Latency (AGL) and Average Routing Overhead Latency (AROL), to establish a more comprehensive assessment mechanism.

\begin{figure*}[t]
  \centering
  % \vspace{-.15in}
  \includegraphics[width=0.95\textwidth]{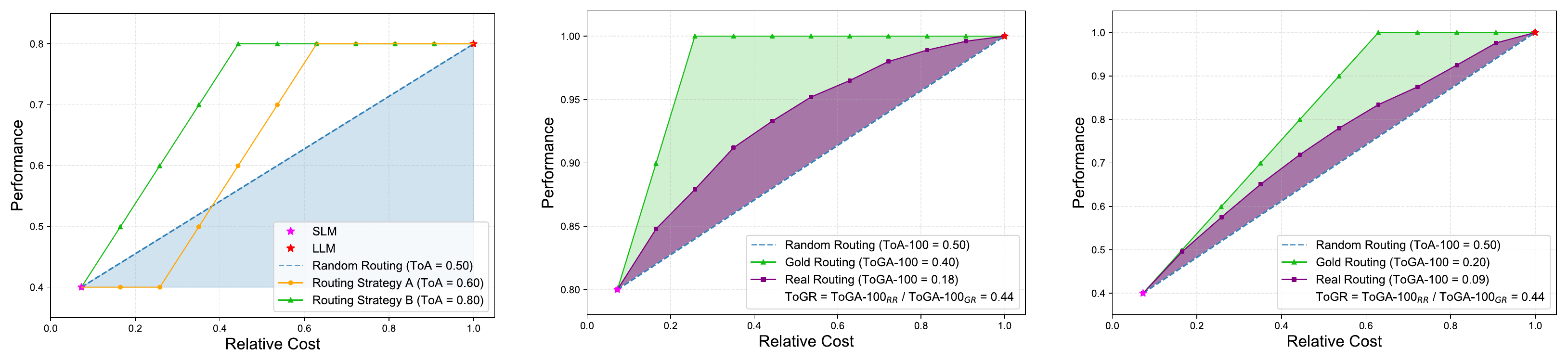}
  % \vspace{-.3in}
  \caption{Introduction to Routing Strategies and Metrics. Strategy A routes the hardest questions (beyond LLM's capability) to LLM first, while Strategy B only routes questions the SLM cannot solve but the LLM can.}
  \label{fig:metrics}
% \vspace{-.15in}
\end{figure*}

To enhance the performance of existing routing strategies and address their limitations, we introduce SATER, a two-stage training approach. In the first stage, shortest-response preference optimization reduces redundant tokens by over 50\% with minimal performance degradation. In the second stage, confidence-based refusal-aware tuning \citep{zhang2023wisdom,cheng2024can,zhang2024r} empowers SLMs to proactively reject complex queries based on confidence thresholds, significantly reducing invalid outputs and latency in cascade routing. Refusal instructions based on different confidence levels can also be approximately applied to pre-generation routing without complex threshold calibration. Evaluations across six widely used benchmarks show that SATER achieves superior ToA and ToGR in pre-generation routing compared to baseline methods. In cascade routing, SATER cuts AGL by over 50\% and AROL by over 80\%, while optimizing cost and accuracy. In summary, our main contributions are as follows:

% SATER provides two key benefits: a) Simple queries are efficiently handled by SLMs without additional threshold calibration, reserving LLMs for complex tasks; b) With hardware support, self-consistency-based voting enables performance comparable to LLMs with minimal latency increase. This seamless integration of methods delivers a flexible, cost-effective solution for LLM applications. In summary, our main contributions are as follows:
\begin{itemize}[noitemsep,topsep=0pt]
\item We formalize a comprehensive evaluation framework to assess routing strategies between small and large language models.
\item We propose SATER, a versatile approach for both pre-generation and cascade routing that shows improved performance while significantly reducing latency.
\item We investigate the comparative advantages of pre-generation and cascade routing under varying conditions, providing practical insights for optimal strategy selection.
\end{itemize}

\section{Problem Formulation}

\subsection{Problem Setting}

\paragraph{Routing Decision Function.} We denote SLM and LLM as \( M_s \) and \( M_l \), respectively, and focus on their routing problem. The routing decision function is then defined as:
\[
r(i) = 
\begin{cases} 
1, & \text{if } s_i < \tau \text{ (routed to } M_l\text{)} \\
0, & \text{if } s_i \geq \tau \text{ (routed to } M_s\text{)}
\end{cases}
\]

where \( s_i \) represents the confidence score predicted for the question \( i \) of \( M_s \), and \( \tau \in [0, 1] \) is the routing threshold. A higher \( \tau \) routes more questions to \( M_l \), improving answer quality at the cost of increased computational resources.

\paragraph{Cost and Performance.} To achieve a more precise evaluation, we perform calculation at the token level. Let \( c_s^{\text{in}} \) and \( c_l^{\text{in}} \) denote the per-token input costs for \( M_s \) and \( M_l \), and \( c_s^{\text{out}} \) and \( c_l^{\text{out}} \) denote their per-token output costs. For a given question \( i \) requiring \( t_i^{\text{in}} \) input tokens, if \( M_s \) generates \( t_i^{s} \) output tokens, its total cost is \( C_i^s = c_s^{\text{in}} \times t_i^{\text{in}} + c_s^{\text{out}} \times t_i^{s} \). Similarly, if \( M_l \) generates \( t_i^{l} \) output tokens, its total cost is \( C_i^l = c_l^{\text{in}} \times t_i^{\text{in}} + c_l^{\text{out}} \times t_i^{l} \).

The total cost for pre-routing is:
\begin{align}
\tilde{C}_{pre} = \frac{\sum_{i=1}^N \left[(1 - r(i))C_i^s + r(i)C_i^l\right]}{\sum_{i=1}^N C_i^l} 
\end{align}

The total cost for cascade routing is:
\begin{align}
\tilde{C}_{cascade} = \frac{\sum_{i=1}^N \left[K \cdot C_i^s + r(i) \cdot C_i^l\right]}{\sum_{i=1}^N C_i^l} 
\end{align}

Here, \( N \) is the total number of questions, and \( K \) is the number of samples. The costs are normalized relative to the cost of using only \( M_l \) (set to 1).We define \( p_i^s \) and \( p_i^l \) as the quality(accuracy) of the answer of \( M_s \) and \( M_l \) for the question \( i \). So, the average quality for both routing methods is:
\begin{align}
\tilde{P} = \frac{1}{N} \sum_{i=1}^N \left[(1 - r(i))p_i^s + r(i)p_i^l\right]
\end{align}

Since \( M_l \) may generate excessively long outputs for complex problems, causing a significant rightward shift in the cost-performance curve if raw token counts are used, thus skewing evaluation results. To address this, we use the average number of output tokens at the dataset level as the per-question cost metric for \( M_l \), while \( M_s \) maintains actual token counts. Additionally, in cascade routing, the KV cache ensures that inputs are computed only once, regardless of the number of samples.

\subsection{Evaluation Metrics}

\paragraph{Cost-Performance Curve.} As shown in Figure~\ref{fig:metrics}, we select threshold points at intervals of 0.1 over \( \tau \) and include two points corresponding to using only the \( M_s \) and \( M_l \): $(C_s, P_s)$ and $(C_l, P_l)$. The curve formed by connecting these points, together with the reference lines $\text{Cost}=C_l$ and $\text{Performance}=P_s$, encloses the \textbf{Trade-off Area (ToA)}, which can be calculated by accumulating trapezoidal areas. The ToA for random routing is 0.5. The \textbf{Trade-off Gain Area (ToGA)} is defined as the ToA improvement over random routing.

\paragraph{ToA-100 and ToGR.} As illustrated in the left part of Figure~\ref{fig:metrics}, the most challenging questions---those that \( M_l \) fails to answer---significantly affect the evaluation results. Although Strategy~B yields superior metric performance, it has notable limitations. First, even when \( M_l \) provides less accurate answers, its responses are generally more informative and insightful than those of \( M_s \). Second, \( M_l \) possesses a higher capability ceiling, and ongoing advancements are likely to address its current shortcomings. Third, a robust routing strategy should not artificially exclude difficult questions to inflate metrics. Therefore, we propose that Strategy~A is the more principled choice. To facilitate a fairer evaluation, we extend ToA and ToGA by introducing \textbf{ToA-100} and \textbf{ToGA-100}, which assume perfect performance of \( M_l \) on all questions.

Moreover, as shown in Figure~\ref{fig:metrics} (middle and right), traditional cost-performance metrics (such as APGR and CPT in RouteLLM\citep{ong2024routellm}) are prone to performance gap bias: when the performance difference between \( M_s \) and \( M_l \) is narrow on a benchmark with easily distinguishable question types or difficulties, routing just a few questions can create the illusion that ``low cost approximates high performance.'' This leads to an overestimation of the router's performance, masking cases where many simple questions are misrouted, while also affecting cross-benchmark evaluation. To mitigate this, we propose the \textbf{Tradeoff Gain Ratio (ToGR)}, which enables fair comparisons by calculating the ratio of ToGA-100 between the current routing and the golden routing. Although the current metric assumes perfect performance of \( M_l \) on all questions, our evaluation can naturally extend to alternative scenarios. While we generally believe that harder questions should be routed to the large model to obtain better answers, from a purely cost-saving perspective, it is actually cheaper not to route these questions at all. Our evaluation framework easily accommodates this case: one only needs to compute ToGR using ToGA instead of ToGA-100.

\paragraph{Latency.} In cascade routing, latency should be a key consideration. (It is important to note that, in this paper, we define cascade routing as requiring the small model to generate a complete output first. This definition may differ from those used in other works.) In non-long-text scenarios, the time required for the generation phase typically far exceeds that of the prefill phase. Moreover, to eliminate the impact of hardware differences, we adopt the number of output tokens as a proxy for latency and define two distinct metrics: 1) \textbf{Average Generation Latency (AGL)}: This represents the latency incurred when the response is ultimately completed by \( M_s \). 2) \textbf{Average Routing Overhead Latency (AROL)}: This quantifies the extra latency when \( M_s \) fails and the system must fall back to \( M_l \), compared to calling \( M_l \) directly. The AGL reflects the processing efficiency of \( M_s \), while the AROL measures the overhead of routing switches. During parallel sampling, early stopping is triggered when the threshold is either exceeded or cannot be reached. Otherwise, the system should default to the longest sample, as it needs to wait for all sampling to complete before making final decisions.

\section{Methodology}

\paragraph{Stage I: Long to Short Training.} Although many previous studies have reported that Direct Preference Optimization (DPO) \citep{rafailov2023direct} struggles to reduce output length in reasoning models, our experiments demonstrate its strong effectiveness in non-reasoning models. Specifically, we integrate standard SFT loss $\mathcal{L}_{\text{SFT}}$ with DPO loss $\mathcal{L}_{\text{DPO}}$, where the latter learns shortest-response preference while the former stabilizes training. The total loss is:

% \[
% \mathcal{L}_{\text{Total}} = \mathcal{L}_{\text{DPO}} + \lambda \mathcal{L}_{\text{SFT}}
% \]
\begin{align}
\mathcal{L}_{\text{Total}} = \mathcal{L}_{\text{DPO}} + \lambda \mathcal{L}_{\text{SFT}}
\end{align}

where $\mathcal{L}_{\text{DPO}}$ is defined as:
% \begin{small}
% \[-\mathbb{E}_{(x, y_w, y_l) \sim \mathcal{D}} \left[ \log \sigma \left( \beta \log \frac{\pi_{\theta}(y_w | x)}{\pi_{\text{ref}}(y_w | x)} - \beta \log \frac{\pi_{\theta}(y_l | x)}{\pi_{\text{ref}}(y_l | x)} \right) \right]\]
% \end{small}

% \begin{multline}
% -\mathbb{E}_{(x, y_w, y_l) \sim \mathcal{D}} \bigg[ 
%    \log \sigma \Big( 
%       \beta \log \frac{\pi_{\theta}(y_w \mid x)}{\pi_{\text{ref}}(y_w \mid x)} 
%       - {} \\
%       \beta \log \frac{\pi_{\theta}(y_l \mid x)}{\pi_{\text{ref}}(y_l \mid x)}
%    \Big) 
% \bigg]
% \end{multline}

\begin{align}
-\mathbb{E}_{(x, y_w, y_l) \sim \mathcal{D}} \bigg[ 
   \log \sigma \Big( 
      & \beta \log \frac{\pi_{\theta}(y_w \mid x)}{\pi_{\text{ref}}(y_w \mid x)} \notag \\
      &- \beta \log \frac{\pi_{\theta}(y_l \mid x)}{\pi_{\text{ref}}(y_l \mid x)}
   \Big) 
\bigg]
\end{align}

where $x$ is the input text, $y_w$ and $y_l$ are positive and negative responses, $\pi_{\theta}$ is the policy model, $\pi_{\text{ref}}$ is the reference model, $\beta$ is the temperature coefficient, and $\sigma$ is the sigmoid function.

To construct training data, we sample each question ten times, selecting the shortest correct response as the positive example and the longest incorrect response, exceeding $1.5$ times the length of the positive sample, as the negative example. This method trains the model to differentiate between concise, correct responses and verbose, incorrect ones. Experiments reveal that DPO is highly sensitive to response length: if $\beta$ and $\lambda$ are set too low, the model generates overly short, low-quality outputs. Thus, we set $\beta = 1$ and $\lambda = 0.2$ to ensure stable training. Additionally, including the longest correct response as a negative example reduces average accuracy by over $2$\%.

\begin{figure*}[t]
  \centering
  % \vspace{-.15in}
  \includegraphics[width=0.95\textwidth]{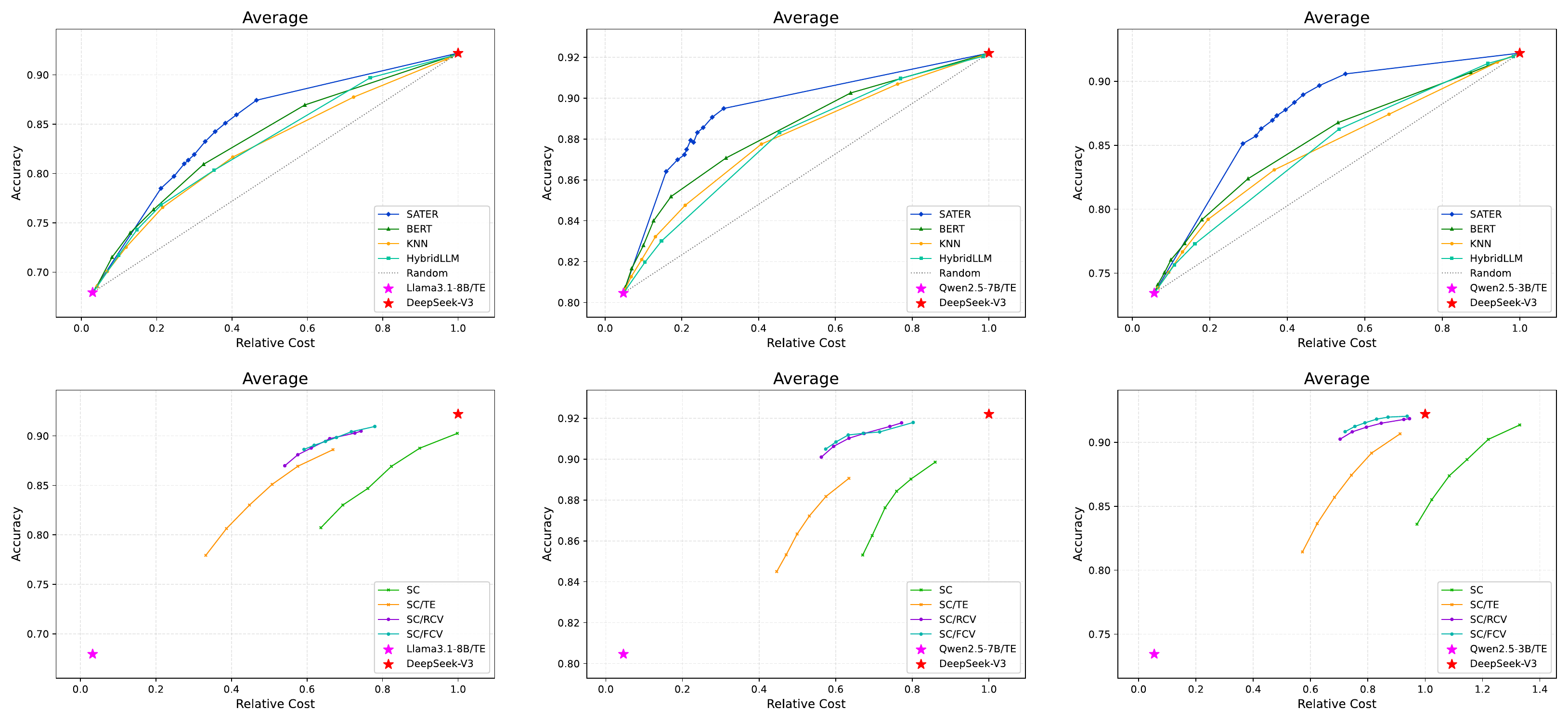}
  % \vspace{-.3in}
  \caption{Average Cost-Accuracy Plot. Results are based on the average of six benchmarks. The top three curves represent pre-generation routing, while the bottom three display cascade routing (cost ratio: 1:13.75). The Average Cost-Accuracy(100) Plot and the individual results for each benchmark are presented in Appendix \ref{sec:avg-cost-accuracy}.}
  \label{fig:main}
% \vspace{-.15in}
\end{figure*}

\paragraph{Stage II: Refusal Training.}  We first use the model trained in Stage I to resample each question ten times, computing its accuracy on a scale from $0$ to $1.0$. Next, we define ten confidence thresholds ranging from $0.1$ to $1.0$. For each question and threshold, we generate new training samples by prepending the prompt: ``Please respond with a confidence level of [threshold]:''. If the question’s accuracy exceeds the threshold, we randomly select a correct answer; otherwise, we apply a rejection template: ``Sorry, I can't answer that.'' Using this training set, we employ the standard SFT loss $\mathcal{L}_{\text{SFT}}$ to fine-tune the model, enabling it to adjust its responses based on confidence levels.

This method can be flexibly applied to various routing strategies. In pre-generation routing, questions rejected by $M_s$ are automatically redirected to $M_l$. For cascade routing, we propose a confidence-based dynamic weighted voting mechanism. For a given question $i$, with $K$ total votes where the $k$-th answer is $a_k \in A$ ($A = \{A_1, \dots, A_M\}$), each discretized confidence score $p_k \in \{0.1, 0.2, \dots, 1.0\}$ is assigned a weight \( w_k = 0.55 + \alpha(p_k - 0.55) \), where $0.55$ represents the average confidence and $\alpha = 0.5$ is a coefficient that ensures higher-confidence answers receive greater weight while mitigating decision bias from individual high-confidence errors. The final confidence score for candidate answer $A_m$ is:
\begin{align} 
\delta(A_m) = \frac{\sum_{k=1}^K w_k \cdot \mathbb{I}(a_k = A_m)}{\sum_{k=1}^K w_k}
\end{align}
and the answer with the highest score is selected. Based on this, we propose two voting schemes: \textbf{Ranged Confidence Voting (RCV)}, which samples confidences uniformly from $0.1$ to $1.0$ for $10$ times, and \textbf{Fixed Confidence Voting (FCV)}, which samples only at a confidence of $1.0$ for $10$ times.

In a pure LLM scenario, refusal training can lead to over-rejection, significantly reducing usability. In contrast, over-rejection in routing systems differs fundamentally: First, it increases computational costs without compromising system availability, as all questions ultimately receive answers. Second, in pre-generation routing, over-rejection resembles an imperfect classifier, whereas in cascade routing, avoiding repeated sampling of unsolvable questions offsets additional costs to a certain extent. 

% Furthermore, we find that models not only identify knowledge gaps and reject answering but also exhibit a robust ability to predict question difficulty in reasoning tasks, proactively rejecting them when appropriate.

\section{Experiments}

\begin{table*}[!ht]
\centering
% \small
\large
\resizebox{\textwidth}{!}{

\begin{tabular}{ll *{12}{c}}
\toprule
\midrule
\multirow{2}{*}{\textbf{Model}} & \multirow{2}{*}{\textbf{Method}} & \multicolumn{2}{c}{\textbf{MMLU}} & \multicolumn{2}{c}{\textbf{MATH-500}} & \multicolumn{2}{c}{\textbf{GSM8K}} & \multicolumn{2}{c}{\textbf{ARC\_C}} & \multicolumn{2}{c}{\textbf{ReClor}} & \multicolumn{2}{c}{\textbf{ARC\_E}} \\
\cmidrule(lr){3-4} \cmidrule(lr){5-6} \cmidrule(lr){7-8} \cmidrule(lr){9-10} \cmidrule(lr){11-12} \cmidrule(lr){13-14}
& & \textbf{ToA-100} & \textbf{ToGR} & \textbf{ToA-100} & \textbf{ToGR} & \textbf{ToA-100} & \textbf{ToGR} & \textbf{ToA-100} & \textbf{ToGR} & \textbf{ToA-100} & \textbf{ToGR} & \textbf{ToA-100} & \textbf{ToGR} \\
\midrule
\multirow{4}{*}{\makecell{Llama-3.1-\\8B-Instruct}} 
& HybridLLM & 0.604 & 0.313 & 0.504 & 0.018 & 0.503 & 0.007 & 0.550 & 0.125 & 0.532 & 0.111 & 0.527 & 0.062 \\
& KNN & 0.596 & 0.290 & 0.583 & 0.415 & 0.519 & 0.050 & 0.542 & 0.107 & 0.522 & 0.076 & 0.518 & 0.041 \\
& BERT & 0.618 & 0.358 & 0.606 & 0.529 & 0.618 & 0.308 & 0.569 & 0.173 & 0.554 & 0.190 & 0.535 & 0.079 \\
& SATER & \textbf{0.655} & \textbf{0.469} & \textbf{0.622} & \textbf{0.607} & \textbf{0.643} & \textbf{0.373} & \textbf{0.702} & \textbf{0.512} & \textbf{0.594} & \textbf{0.326} & \textbf{0.667} & \textbf{0.379} \\
\midrule
\multirow{4}{*}{\makecell{Qwen2.5-\\7B-Instruct}} 
& HybridLLM & 0.586 & 0.239 & 0.502 & 0.006 & 0.494 & $-$0.014 & 0.465 & $-$0.080 & 0.512 & 0.032 & 0.455 & $-$0.098 \\
& KNN & 0.619 & 0.332 & 0.666 & 0.498 & 0.576 & 0.168 & 0.512 & 0.026 & 0.467 & $-$0.090 & 0.481 & $-$0.042 \\
& BERT & 0.643 & 0.396 & 0.702 & 0.605 & 0.650 & 0.334 & 0.566 & 0.148 & 0.545 & 0.124 & 0.544 & 0.095 \\
& SATER & \textbf{0.692} & \textbf{0.533} & \textbf{0.770} & \textbf{0.810} & \textbf{0.708} & \textbf{0.461} & \textbf{0.631} & \textbf{0.294} & \textbf{0.639} & \textbf{0.385} & \textbf{0.593} & \textbf{0.201} \\
\midrule
\multirow{4}{*}{\makecell{Qwen2.5-\\3B-Instruct}} 
& HybridLLM & 0.618 & 0.352 & 0.512 & 0.042 & 0.483 & $-$0.043 & 0.497 & $-$0.006 & 0.512 & 0.039 & 0.486 & $-$0.030 \\
& KNN & 0.620 & 0.358 & 0.648 & 0.520 & 0.538 & 0.096 & 0.493 & $-$0.016 & 0.509 & 0.028 & 0.492 & $-$0.017 \\
& BERT & 0.630 & 0.389 & 0.660 & 0.562 & 0.641 & 0.361 & 0.537 & 0.090 & 0.528 & 0.087 & 0.569 & 0.151 \\
& SATER & \textbf{0.687} & \textbf{0.560} & \textbf{0.711} & \textbf{0.740} & \textbf{0.741} & \textbf{0.615} & \textbf{0.685} & \textbf{0.445} & \textbf{0.600} & \textbf{0.311} & \textbf{0.639} & \textbf{0.307} \\
\midrule
\bottomrule
\end{tabular}}

\caption{ToA-100 and ToGR results across in-domain and out-of domain datasets. Bold indicates the best.}
\label{tab: togr}
\end{table*}

\begin{table*}[!ht]
\centering
\footnotesize
\resizebox{\textwidth}{!}{
% \adjustbox{max width=\textwidth}{
\begin{tabular}{ll *{14}{c}}
\toprule
\midrule
\multirow{2}{*}{\textbf{Model}} & \multirow{2}{*}{\textbf{Method}} & \multicolumn{2}{c}{\textbf{MMLU}} & \multicolumn{2}{c}{\textbf{MATH-500}} & \multicolumn{2}{c}{\textbf{GSM8K}} & \multicolumn{2}{c}{\textbf{ARC\_C}} & \multicolumn{2}{c}{\textbf{ReClor}} & \multicolumn{2}{c}{\textbf{ARC\_E}} & \multicolumn{2}{c}{\textbf{Average}} \\
\cmidrule(lr){3-4} \cmidrule(lr){5-6} \cmidrule(lr){7-8} \cmidrule(lr){9-10} \cmidrule(lr){11-12} \cmidrule(lr){13-14} \cmidrule(lr){15-16}
& & \textbf{AGL} & \textbf{AROL} & \textbf{AGL} & \textbf{AROL} & \textbf{AGL} & \textbf{AROL} & \textbf{AGL} & \textbf{AROL} & \textbf{AGL} & \textbf{AROL} & \textbf{AGL} & \textbf{AROL} & \textbf{AGL} & \textbf{AROL}\\
\midrule
\multirow{4}{*}{\makecell{Llama-3.1-\\8B-Instruct}} & SC & 186 & 293 & 365 & 638 & 226 & 306 & 140 & 216 & 150 & 177 & 128 & 195 & 199 & 304
\\
& SC/TE & 68 & 100 & 173 & 211 & 139 & 133 & 49 & 58 & 90 & 90 & 38 & 72 & 93 & 111 \\
& SC/RCV & \underline{48} & \underline{6} & \underline{142} & \underline{31} & \underline{133} & \underline{75} & \underline{40} & \underline{4} & \underline{76} & \underline{4} & \underline{30} & \underline{2} & \underline{78} & \underline{20}
\\
& SC/FCV & \textbf{47} & \textbf{1} & \textbf{125} & \textbf{7} & \textbf{126} & \textbf{22} & \textbf{38} & \textbf{1} & \textbf{74} & \textbf{2} & \textbf{29} & \textbf{1} & \textbf{73} & \textbf{6}
\\
\midrule
\multirow{4}{*}{\makecell{Qwen2.5-\\7B-Instruct}} & SC & 182 & 437 & 477 & 825 & 342 & 432 & 126 & 227 & 335 & 421 & 98 & 189 & 260 & 422
\\
& SC/TE & 114 & 240 & 361 & 533 & 271 & 313 & \textbf{75} & 130 & 172 & 262 & \textbf{60} & 142 & 176 & 270
\\
& SC/RCV & \underline{104} & \underline{12} & \underline{310} & \underline{83} & \underline{250} & \underline{108} & 78 & \underline{7} & \underline{142} & \underline{17} & 64 & \underline{6} & \underline{158} & \underline{39} 
\\
& SC/FCV & \textbf{96} & \textbf{4} & \textbf{291} & \textbf{18} & \textbf{243} & \textbf{37} & \underline{77} & \textbf{6} & \textbf{133} & \textbf{10} & \underline{63} & \textbf{4} & \textbf{150} & \textbf{13} 
\\
\midrule
\multirow{4}{*}{\makecell{Qwen2.5-\\3B-Instruct}} & SC & 281 & 425 & 481 & 794 & 369 & 425 & 231 & 341 & 432 & 483 & 188 & 327 & 330 & 466
\\
& SC/TE & 148 & 272 & 346 & 504 & 251 & 273 & 101 & 166 & 287 & 345 & 74 & 146 & 201 & 284
\\
& SC/RCV & \underline{99} & \underline{9} & \underline{284} & \underline{32} & \underline{225} & \underline{71} & \underline{84} & \underline{12} & \underline{233} & \underline{21} & \underline{62} & \underline{10} & \underline{164} & \underline{26} 
\\
& SC/FCV & \textbf{87} & \textbf{2} & \textbf{255} & \textbf{3} & \textbf{207} & \textbf{12} & \textbf{78} & \textbf{4} & \textbf{214} & \textbf{3} & \textbf{60} & \textbf{4} & \textbf{150} & \textbf{5}
\\
\midrule
\bottomrule
\end{tabular}
}
\caption{AGL and AROL results with threshold $\tau = 0.6$. Results with $\tau = 1.0$ are provided in Table \ref{tab: agl-arol-threshold-1}.}
\label{tab: arol}
\end{table*}

\subsection{Experiment Setup} 

\paragraph{Models.} We conduct experiments on three SLMs: Llama-3.1-8B-Instruct \citep{grattafiori2024llama}, Qwen2.5-7B-Instruct, and Qwen2.5-3B-Instruct \citep{yang2024qwen2}, as well as one LLM: DeepSeek-V3-0324 \citep{liu2024deepseek}, a highly cost-effective and high-performing model. To simplify cost comparisons amid opaque API pricing, we adopt Groq’s pricing for SLMs at \$0.08 per million output tokens and DeepSeek’s official rate for the LLM at \$1.10. Input pricing is set at one-quarter of the respective output price, resulting in a cost ratio of 1:13.75. 

% Since SLMs are typically deployed privately (which is also a basic requirement of our method), their actual costs can be significantly lower than \$0.08. In contrast, LLMs generally rely on API calls, with fixed pricing and the potential to use models more expensive than DeepSeek-V3. To account for this, we further explore scenarios with SLM-to-LLM price ratios of 1:25, 1:50, and 1:100.

\paragraph{Datasets.} We conduct experiments on six widely used datasets, covering various types of tasks such as knowledge-based question answering, scientific reasoning, mathematical reasoning, and logical reasoning. These include: MMLU \citep{hendrycks2020measuring}, ARC-Challenge and ARC-Easy \citep{clark2018think}, GSM8K \citep{cobbe2021training}, MATH-500 \citep{hendrycks2021measuring}, and Reclor \citep{yu2020reclor}. For MMLU, lacking a training set, we use its 14,042-question test set for training and 1,531-question validation set for testing. For ReClor, with unavailable test set answers, we use its 500-question validation set for testing. The models are trained on the training sets of MMLU, ARC-Challenge, GSM8K, and MATH-500, with their test sets used for in-domain evaluation. ARC-Easy and ReClor served as out-of-domain datasets.

\paragraph{Baselines.} In pre-generation routing, we compare three mainstream approaches: a BERT-based classifier \citep{ong2024routellm}, a KNN classifier \citep{hu2024routerbench}, and HybridLLM \citep{dinghybrid}, which trains DeBERTa-v3-large (300M) \citep{he2020deberta} using soft labels derived from BART scores and multiple sampling. As routing modules often operate in resource-constrained environments, we avoid larger classification models, which are impractical and directly using a larger SLM, such as Qwen2.5-14B-Instruct is generally more effective. In cascade routing, we compare the original model (SC), the model trained only in stage I (SC/TE), and the model after full two-stage training with two voting methods (RCV, FCV). For more implementation details, please refer to Appendix \ref{sec: implementation-details}.

% We train the BERT classifier and HybridLLM for 5 epochs, selecting the best-performing epoch on the test set for evaluation. KNN routing uses RouterBench’s recommended configuration (40 nearest neighbors, all-MiniLM-L12-v2 embedding model). 

\subsection{Main Results}

\paragraph{SATER demonstrates superior performance and adaptability in pre-generation routing.} Experiments reveal that SATER consistently outperforms baseline methods across three SLMs and six datasets. As illustrated in Figure \ref{fig:main} and \ref{fig:main_100}, SATER achieves significantly higher average performance, with Figure \ref{fig:qwen_1375}, \ref{fig:qwen_vote_1375} and Table \ref{tab: togr} confirming stable advantages in all single-dataset evaluations. Notably, SATER exhibits strong task adaptability and generalizability: It automatically adjusts routing intervals based on task complexity and performs well in out-of-distribution tests, achieving ToGR scores averaging 0.2 higher than the BERT classifier on both the complex ReClor and simpler ARC\_E datasets.

\begin{figure*}[t]
  \centering
  % \vspace{-.15in}
  \includegraphics[width=0.95\textwidth]{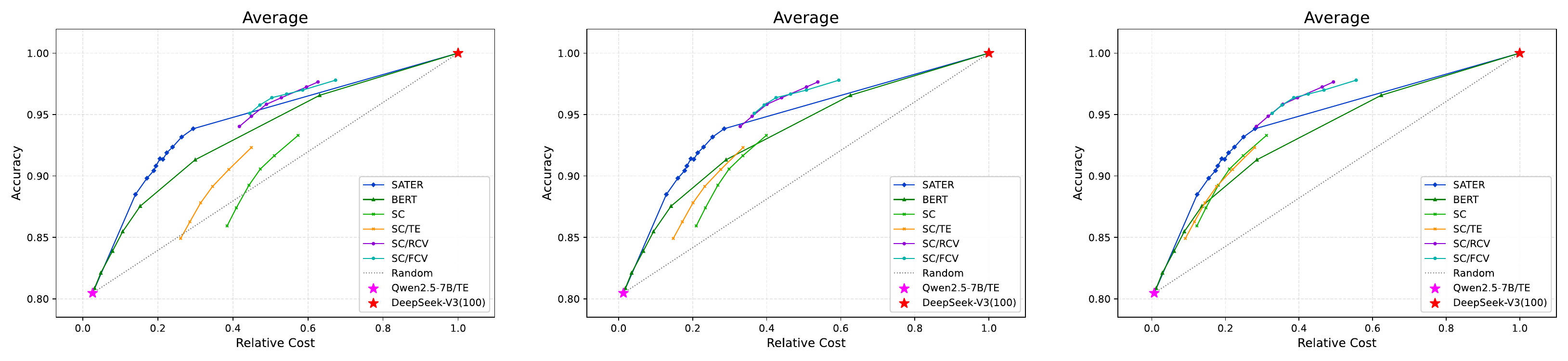}
  % \vspace{-.3in}
  \caption{Comparison plot of cost-accuracy(100) between pre-generation and cascade routing, averaged across all benchmarks. Results are based on Qwen2.5-7B-Instruct, with three subplots depicting cost ratios of 1:25, 1:50, and 1:100 from left to right. Detailed results for individual benchmarks are available in Appendix \ref{sec:comparison-plot} (Figure \ref{fig:rv_25}, \ref{fig:rv_50}, \ref{fig:rv_100}).}
  \label{fig:rv_25_50_100}
% \vspace{-.10in}
\end{figure*}

\paragraph{ToGR serves as a superior metric, underscoring the necessity of fine-grained task difficulty differentiation in pre-generation routing.} As shown in Figure \ref{fig:qwen_1375} and Table \ref{tab: togr}, Llama-3.1-8B-Instruct scores higher on ToA-100 for datasets with smaller performance gaps, such as GSM8K and ARC, than those with larger gaps, such as MATH and MMLU. However, ToGR exhibits an opposite trend, revealing a performance gap bias and proving a more robust metric. Although all methods struggle with small-gap datasets, necessitating further improvement, SATER significantly outperforms baselines, suggesting that SATER not only enables coarse-grained domain classification, but also possesses finer-grained difficulty discrimination capabilities. 

\paragraph{Models can assess question difficulty and refuse to answer, even in reasoning tasks.} ToGR in Table \ref{tab: togr} reveals an interesting phenomenon: Model performance on reasoning tasks is comparable to knowledge-based tasks, with all models achieving best results on MATH-500. This suggests that models can not only identify the knowledge blind spot, but also anticipate question difficulty in reasoning tasks and proactively decline to answer.

\paragraph{SATER continuously reduces latency and cost while improving performance in cascade routing.} As presented in Tables \ref{tab: arol}, both voting methods significantly reduce latency, with AGL decreasing by over 50\% and AROL by over 80\%. Furthermore, Figure \ref{fig:main} and \ref{fig:qwen_vote_1375} illustrate that SATER surpasses vanilla SC in both cost and accuracy. For stronger SLMs, such as Qwen2.5-7B-Instruct, RCV is more effective in lowering costs by increasing voting opportunities. For weaker SLMs, FCV improves efficiency by swiftly declining to respond. In particular, when LLMs are assumed to deliver optimal responses, the advantages of SATER become even more evident, indicating that the overconfidence of vanilla SC is partially masking by the insufficiency of LLM.

% Furthermore, we find that models not only identify knowledge gaps and reject answering but also exhibit a robust ability to predict question difficulty in reasoning tasks, proactively rejecting them when appropriate.

\section{Analysis}

\subsection{Comparison Between Pre-generation Routing and Cascade Routing}
\paragraph{Different Cost Ratios.} Since SLMs are typically deployed privately, which is also a basic requirement of our method, their actual costs can be much lower than \$0.08. In contrast, LLMs generally rely on API calls, with fixed pricing and the potential to use models more expensive than DeepSeek-V3. So we further explore scenarios with cost ratios of 1:25, 1:50, and 1:100 in Figure \ref{fig:rv_25_50_100}. At low cost ratios, cascade routing is less cost-effective than pre-generation routing due to multiple sampling. As cost ratios rise, cascade routing's benefits emerge: RCV and FCV outperform pre-generation routing at a 1:25 ratio, while SC/TE equals BERT classifier at 1:50 and surpasses it at 1:100. Thus, pre-generation routing excels at low cost ratios, but cascade routing offers superior cost control and accuracy at high ratios. Although RCV and FCV's cost advantages diminish at high ratios, they remain more cost-effective than vanilla SC, with significant advantages in latency control and accuracy.
\paragraph{Different Tasks.} Besides overall performance across six benchmarks, we further analyze the suitability of different routing strategies for various types of tasks. As shown in Figure \ref{fig:rv_25}, \ref{fig:rv_50}, and \ref{fig:rv_100}, cascade routing demonstrates a clear advantage in complex reasoning tasks such as mathematical reasoning. In contrast, for knowledge-intensive tasks like factual question answering, direct routing based on question classification exhibits higher reliability, since self-consistency may not adequately reflect answer accuracy. Notably, SATER consistently outperforms vanilla SC in one or more aspects of cost, accuracy, and latency across all experimental scenarios.

\paragraph{Different Capabilities.} Figure \ref{fig:dk_13_50_100} illustrates that under the same cost ratio, the weaker the capabilities of an LLM, the sooner the advantages of cascade routing become evident. This is primarily because, despite overconfidence issues in voting-based routing, some misjudged difficult queries are actually unsolvable by the LLM, in which case no routing saves costs instead. A secondary reason is that the inherent randomness in LLM outputs may lead to incorrect responses for simple queries in a single sampling, whereas the SLM's accurate voting helps narrow the accuracy gap.

\begin{figure*}[t]
  \centering
  % \vspace{-.15in}
  \includegraphics[width=1.0\textwidth]{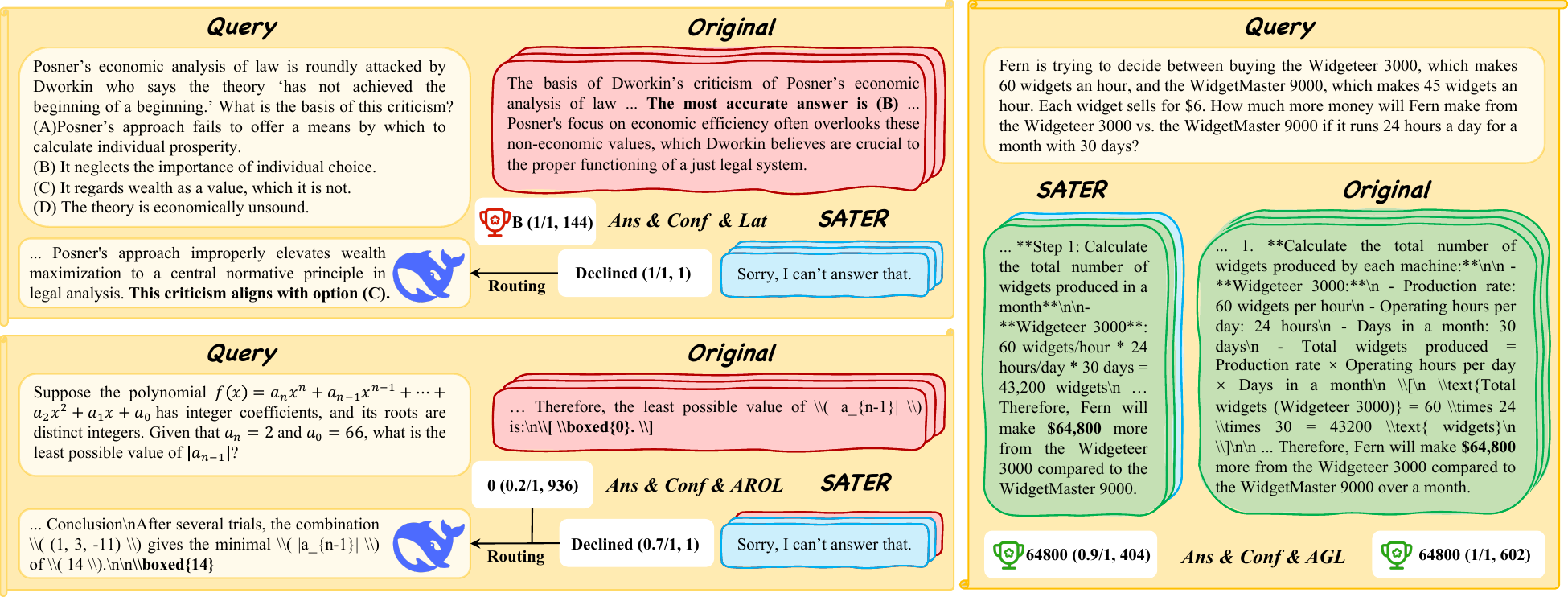}
  % \vspace{-.3in}
  \caption{Three examples from SATER. Responses are color-coded: red (incorrect), green (correct), blue (refused). White box shows the majority-voted answer, confidence score, and AGL or AROL, based on routing decisions. }
  \label{fig:case_study}
% \vspace{-.10in}
\end{figure*}

% \subsection{Case Study: When Does SATER Help?}
\subsection{When Does SATER Help? A Case Study}
In pre-generation routing, SATER resembles embedding a classifier within the model, fully leveraging its parameters. Compared to methods that rely on additional small classifiers, SATER significantly enhances performance under resource-constrained conditions. In addition, the consistency between the model and classifier aids in a fine-grained awareness of capability boundaries. In cascade routing, long-to-short training accelerates the generation process, while refusal training enables the model to decline complex queries. Together, two-stage training not only reduces costs and latency, but also improves accuracy. Figure \ref{fig:case_study} illustrates several examples from MMLU, Math-500, and GSM8K. For overconfident queries, especially knowledge-based ones, SATER firmly rejects them; for challenging reasoning tasks, SATER halts generation with a high rejection probability, reducing ineffective sampling and AROL; for medium and simple queries, SATER swiftly generates responses, further boosting overall efficiency. 

\subsection{Other Experiments}
% Tables \ref{tab: accuracy-tokens} and \ref{tab: delta-accuracy-tokens} in Appendix \ref{sec:lts-effectiveness} show that SATER reduces token count by over 40\% in first-stage DPO training while maintaining high accuracy. Table \ref{tab: efficiency-comparison} compares SATER and TokenSkip on GSM8K and MATH-500, highlighting SATER's superior token compression, accuracy retention, and readability. These results confirm DPO's effectiveness as a simple yet effective method for shortening output while preserving performance. Additionally, Table 3 shows that lowering the sampling temperature reduces output diversity and leads to a decline in accuracy in the high-threshold interval. Among them, SC and RCV, which rely on voting mechanisms, exhibit slightly greater performance degradation compared to FCV, which relies more on a rejection mechanism.Overall, RCV and FCV still maintain a relatively significant advantage.

Table \ref{tab: accuracy-tokens} and \ref{tab: delta-accuracy-tokens} in Appendix \ref{sec:lts-effectiveness} demonstrate that SATER reduces token count by over 40\% in first-stage DPO training while preserving high accuracy. Table \ref{tab: efficiency-comparison} compares SATER and TokenSkip on GSM8K and MATH-500, showing SATER's superior token compression, accuracy retention, and readability. These findings highlight DPO's effectiveness in shortening output without compromising performance. Additionally, Figure \ref{fig:t03} indicates that lower sampling temperatures reduce output diversity, leading to accuracy declines in high-threshold intervals. However, RCV and FCV still maintain a relatively significant advantage.

\section{Related Work}

\paragraph{Pre-generation Routing} Pre-generation routing intelligently assigns tasks by evaluating query features, mainly through two approaches: domain expert routing \citep{stripelis2024tensoropera,lu2024routing,chen2024routerdc}, which identifies the query’s domain and assigns it to a specialized model for improved performance, and complexity-adaptive routing \citep{dinghybrid, ong2024routellm}, which allocates queries to models of varying sizes based on task complexity, optimizing the balance between quality and cost. Both rely on similar frameworks, such as supervised classifiers or unsupervised clustering, but differ in implementation details, including input query types and label annotations.
% hu2024routerbench,vsakota2024fly,chen2024routerdc
% (usually with similar parameter scales and costs) 
\paragraph{Cascade Routing} Cascade routing starts with a smaller model generating a response, evaluated for quality. If it falls short of a set threshold, the query is passed to a stronger model until satisfactory. Decisions use either task-specific evaluation models for simple classification tasks \citep{chen2023frugalgpt, ramirez2024optimising} or confidence scores for more complex tasks\citep{aggarwal2024automix,yuelarge}, which generally require multiple samplings. 

% Compared to pre-generation routing, cascade routing delivers superior performance across various tasks but introduces higher latency.

\paragraph{LLM Honesty} Honesty in LLMs refers to their ability to produce truthful and reliable outputs. Current approaches, including supervised fine-tuning \citep{zhang2024r,cheng2024can}, reinforcement learning \citep{xu2024rejection}, and probing \citep{kossen2024semantic}, encourage models to admit uncertainty by saying ``I don't know'' or giving better confidence estimates. While most studies target knowledge-intensive tasks like factual question answering, our work shows that for complex, multi-step reasoning tasks, models can proactively decline to answer, thereby optimizing two types of routing across diverse tasks.

\paragraph{Efficient Reasoning} Long CoT reasoning models often ``overthink'', producing redundant steps that reduce efficiency. Current training solutions include length-penalized reinforcement learning for concise reasoning \citep{luo2025o1, aggarwal2025l1}, and supervised fine-tuning with compact CoT for succinct outputs \citep{xia2025tokenskip,munkhbat2025self}. Studies like TokenSkip also indicate that output redundancy persists, even in non-reasoning models, though less severely.
% yeo2025demystifying ma2025cot,
% also indicate that output redundancy, while less pronounced in non-reasoning models, remains a significant challenge.

\section{Conclusion}
In this work, we introduce a comprehensive evaluation framework and propose SATER, a simple yet effective two-stage training approach. Extensive experiments across various SLMs and tasks demonstrate SATER's effectiveness, achieving significant improvements in performance, cost efficiency, and latency for both pre-generation routing and cascade routing. When the cost ratio between LLM and SLM exceeds 50, SATER delivers performance comparable to pure LLM at around 50\% of the cost while maintaining low latency. Further analysis highlights the suitability of both routing strategies under diverse conditions, offering a flexible and cost-effective solution for LLM applications.

\section*{Limitation}
Although our work demonstrates strong results, certain limitations remain. First, we primarily focus on the routing mechanism between a single small language model and a single large language model. Future research could explore multi-model collaborative routing or cascading to enhance scalability. Secondly, the prompt-based refusal training struggles to achieve effective routing when the threshold falls below 0.1. However, cross-dataset experiments demonstrate that setting the threshold at 0.1 maintains the overall cost at less than 30\% of the LLM's cost, indicating that SATER consistently delivers a practical range of routing costs. Finally, SATER's rejection mechanism faces challenges in scenarios where a response is mandatory but cannot be routed. In such cases, maintaining an untrained model copy may be necessary to ensure the system can still provide complete responses when required.

% \section*{Acknowledgments}

\bibliography{custom}

% \newpage
\appendix

\section{Additional Plots}

% \subsection{Average cost-accuracy(100) Plot}
% \label{sec:avg-cost-accuracy}

\subsection{Average Cost-Accuracy(100) Plot and Cost-Accuracy(100) Plot for Each Benchmark}
\label{sec:avg-cost-accuracy}
\begin{figure*}[t]
  \centering
  % \vspace{-.15in}
  \includegraphics[width=0.95\textwidth]{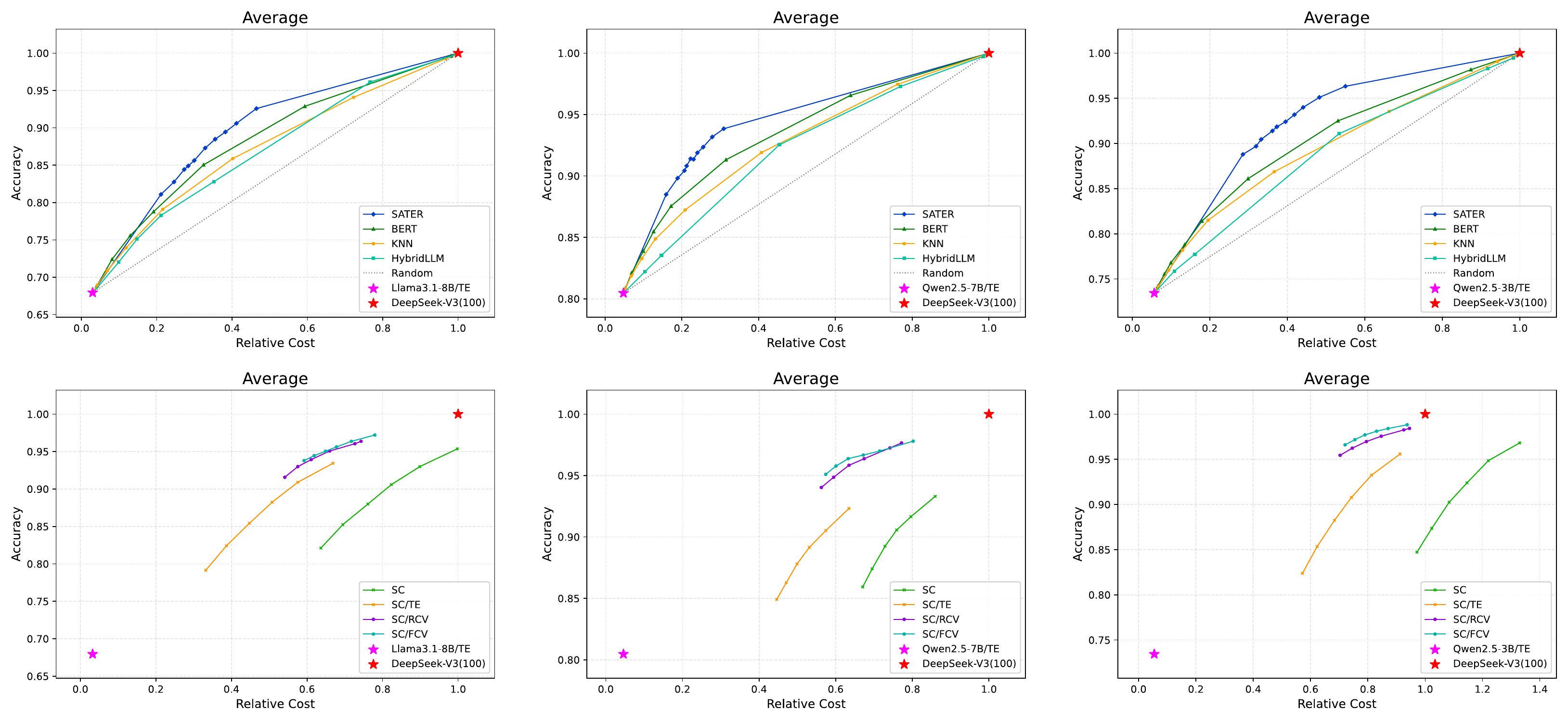}
  % \vspace{-.3in}
  \caption{Average Cost-Accuracy(100) Plot.}
  \label{fig:main_100}
% \vspace{-.15in}
\end{figure*}

\begin{figure*}[t]
  \centering
  % \vspace{-.15in}
  \includegraphics[width=0.95\textwidth]{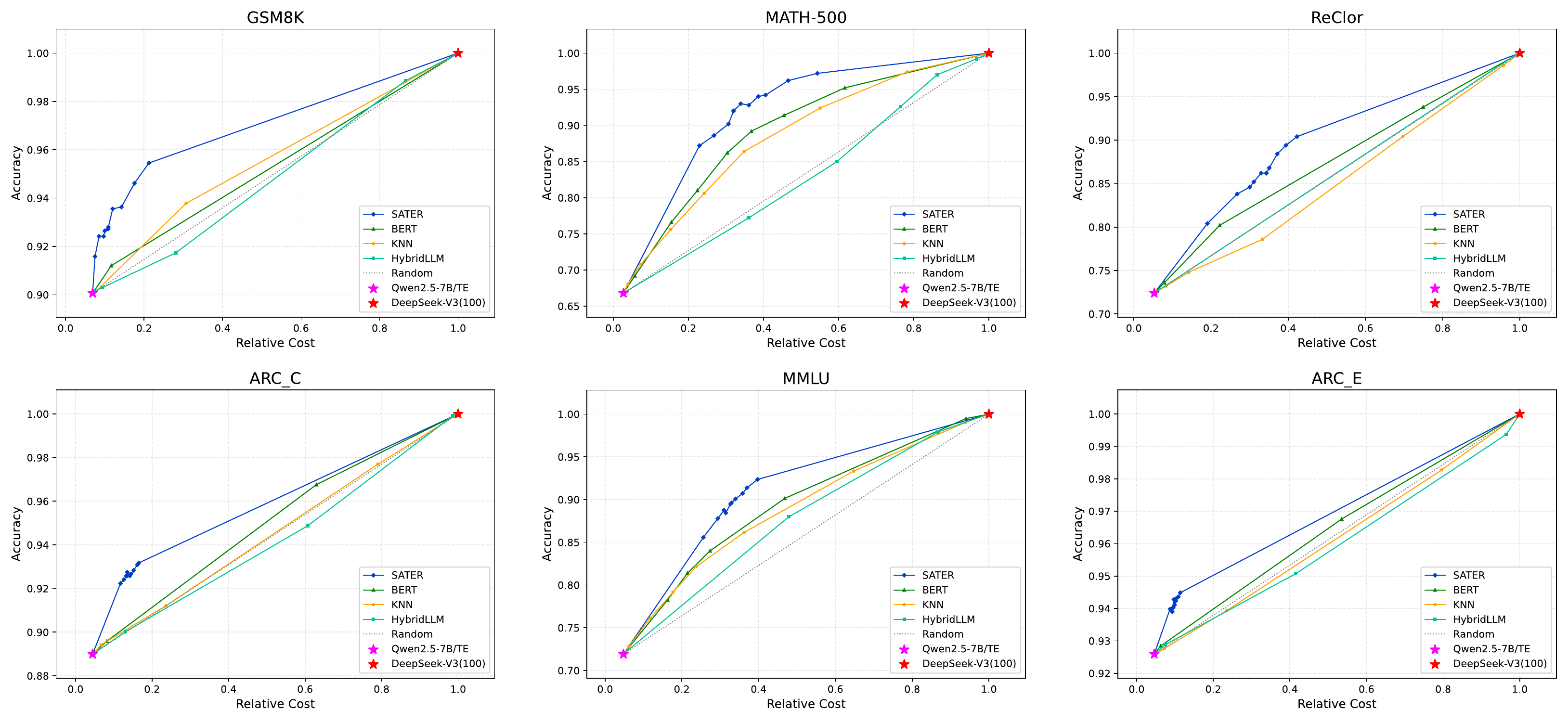}
  % \vspace{-.3in}
  \caption{Cost-Accuracy(100) plot for each benchmark in pre-generation routing. Results are based on Qwen2.5-7B-Instruct, with a cost ratio of 1:13.75.}
  \label{fig:qwen_1375}
% \vspace{-.15in}
\end{figure*}

\begin{figure*}[t]
  \centering
  % \vspace{-.15in}
  \includegraphics[width=0.95\textwidth]{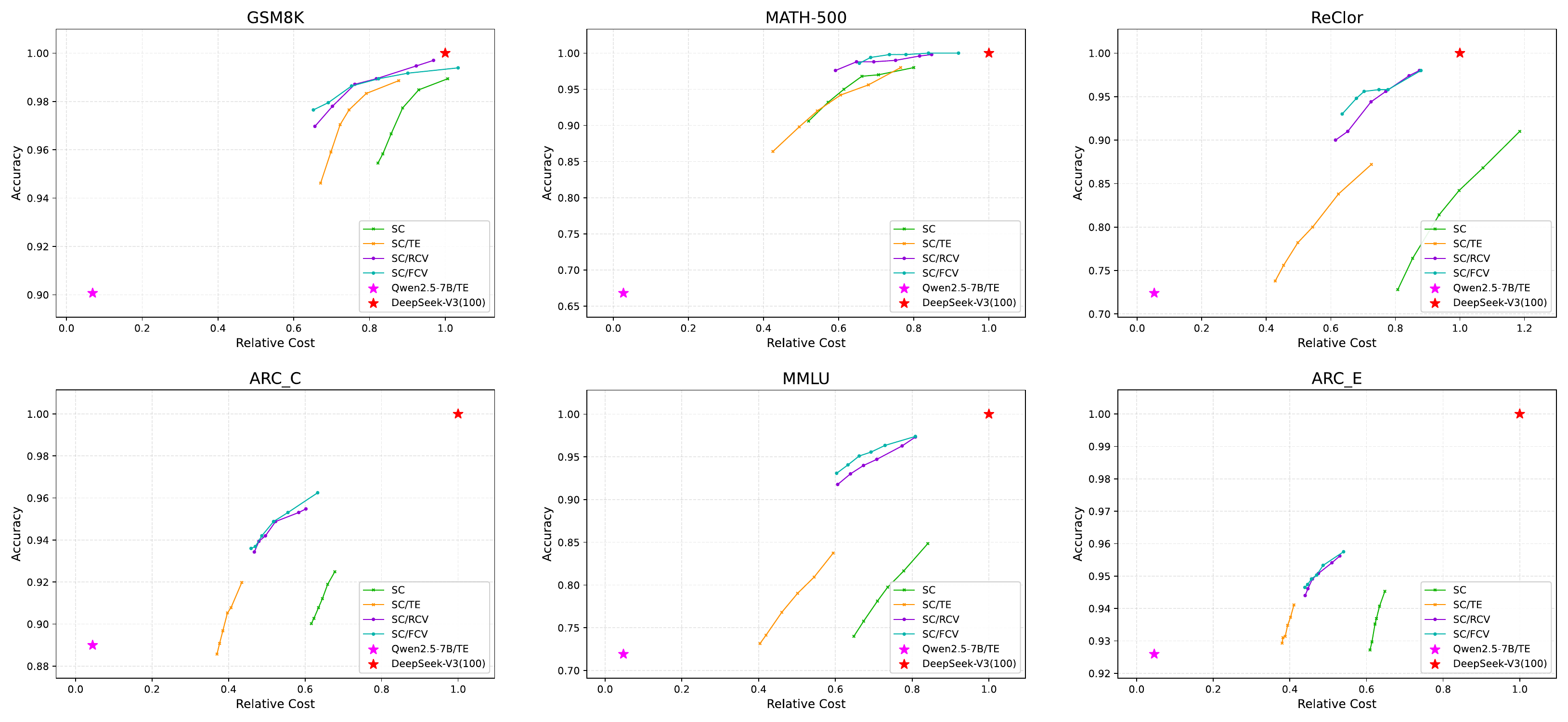}
  % \vspace{-.3in}
  \caption{Cost-Accuracy(100) plot for each benchmark in cascade routing. Results are based on Qwen2.5-7B-Instruct, with a cost ratio of 1:13.75.}
  \label{fig:qwen_vote_1375}
% \vspace{-.15in}
\end{figure*}

Figure \ref{fig:main_100} displays the average cost-accuracy(100) plot across all benchmarks. The detailed cost-accuracy(100) performance for individual benchmarks is shown in Figure \ref{fig:qwen_1375} (pre-generation routing) and Figure \ref{fig:qwen_vote_1375} (cascade routing). All results are obtained using Qwen2.5-7B-Instruct, with a cost ratio of 1:13.75.

\subsection{Comparison Plot of Cost-Accuracy(100) between Pre-generation Routing and Cascade Routing for Each Benchmark}
\label{sec:comparison-plot}
\begin{figure*}[t]
  \centering
  % \vspace{-.15in}
  \includegraphics[width=0.95\textwidth]{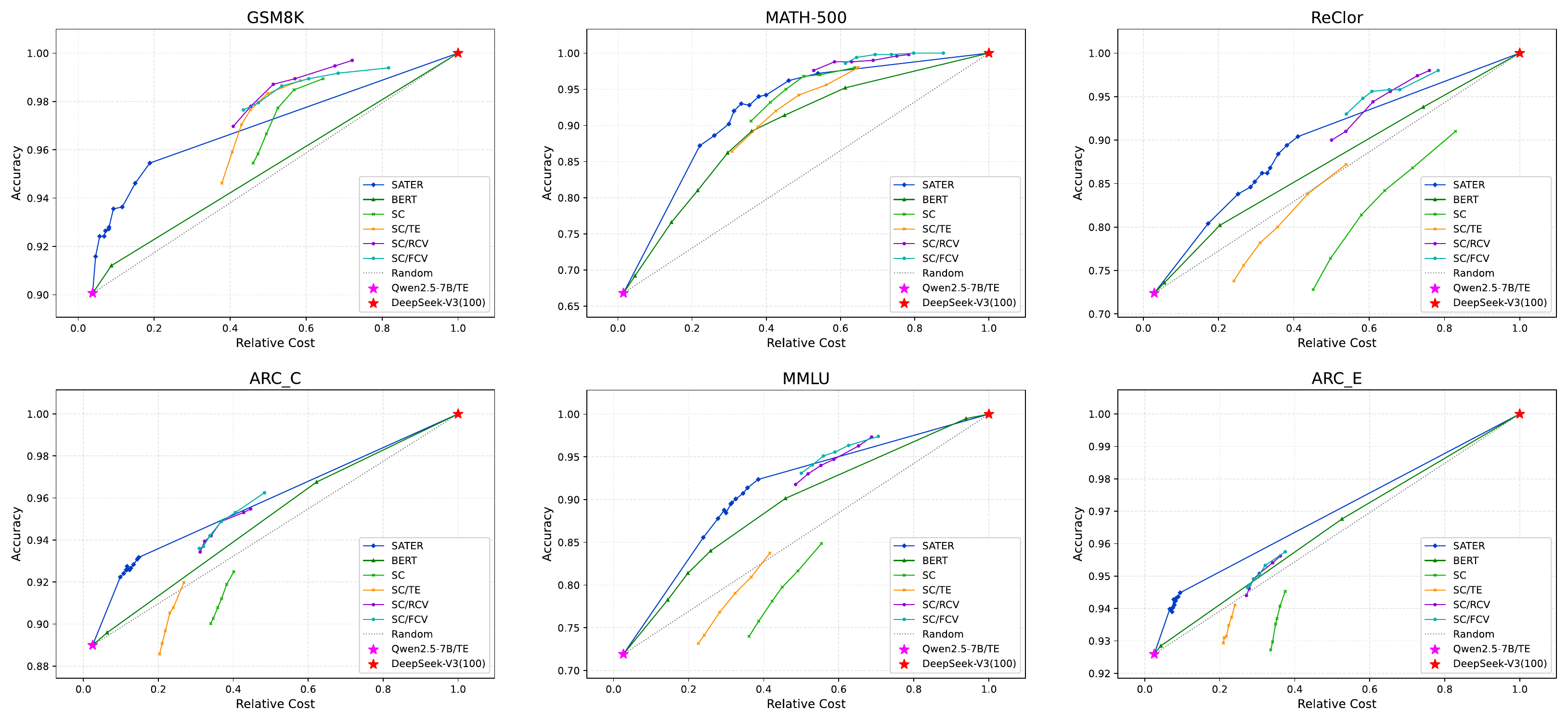}
  % \vspace{-.3in}
  \caption{Comparison plot of cost-accuracy(100) between pre-generation routing and cascade routing for each benchmark. Results are based on Qwen2.5-7B-Instruct, with a cost ratio of 1:25.}
  \label{fig:rv_25}
\vspace{-.10in}
\end{figure*}

\begin{figure*}[t]
  \centering
  % \vspace{-.15in}
  \includegraphics[width=0.95\textwidth]{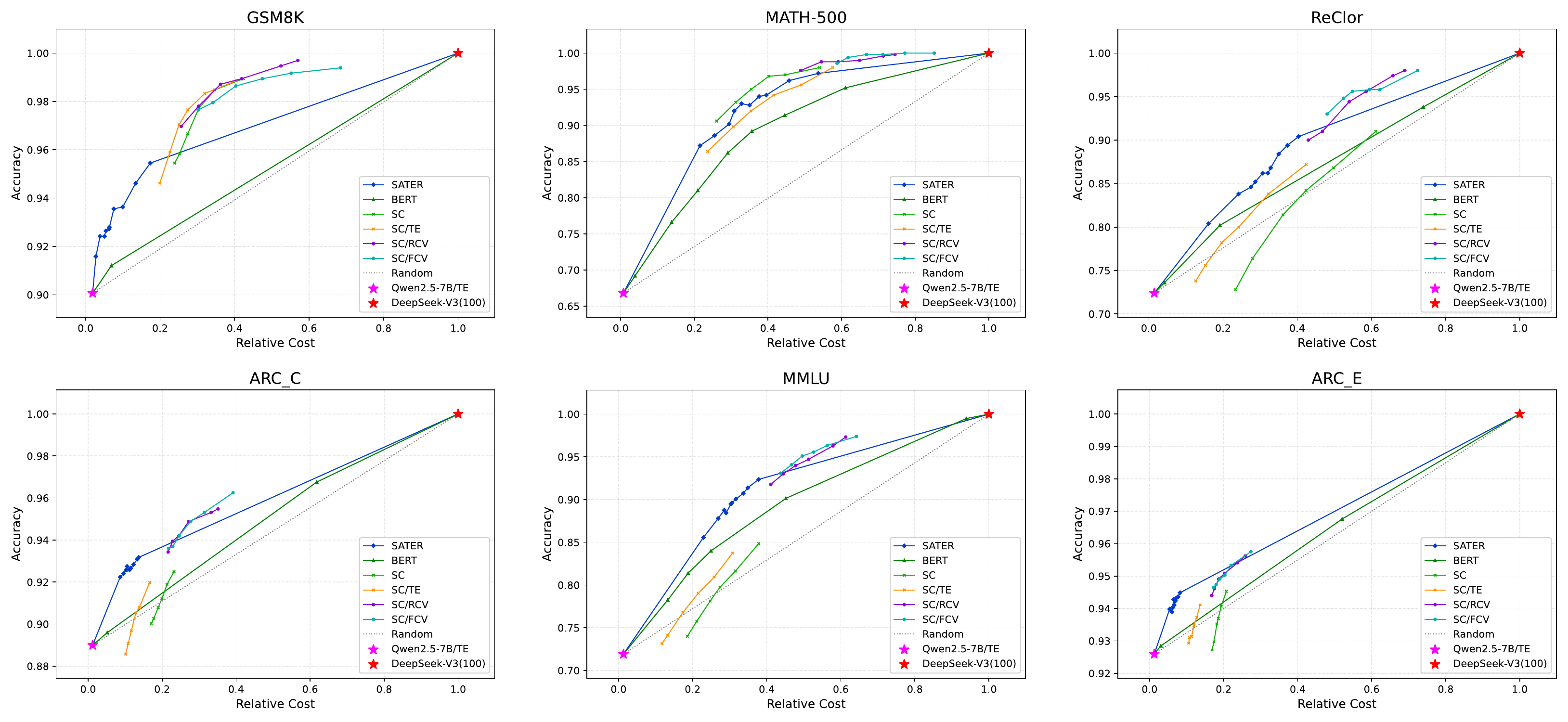}
  % \vspace{-.3in}
  \caption{Comparison plot of cost-accuracy(100) between pre-generation routing and cascade routing for each benchmark. Results are based on Qwen2.5-7B-Instruct, with a cost ratio of 1:50.}
  \label{fig:rv_50}
\vspace{-.10in}
\end{figure*}

\begin{figure*}[t]
  \centering
  % \vspace{-.15in}
  \includegraphics[width=0.95\textwidth]{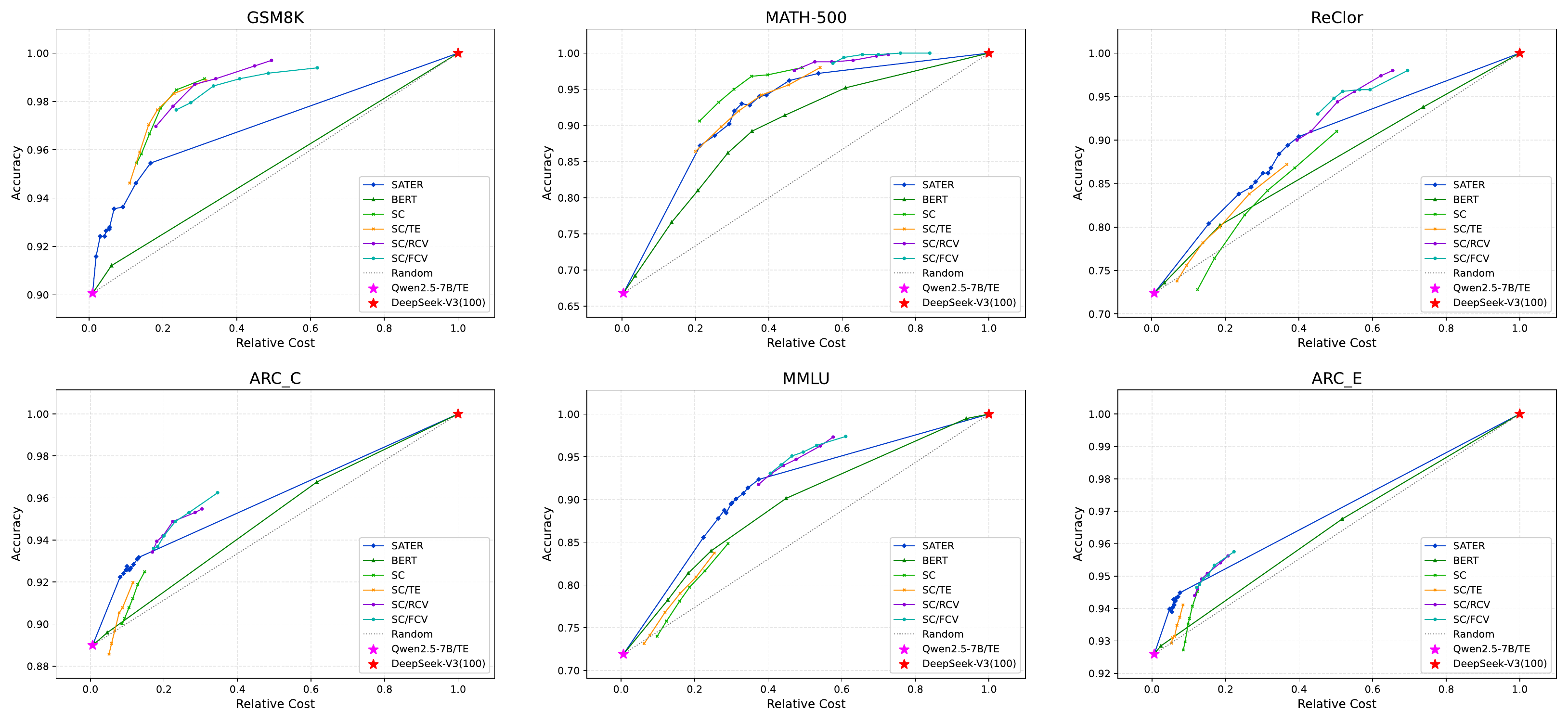}
  % \vspace{-.3in}
  \caption{Comparison plot of cost-accuracy(100) between pre-generation routing and cascade routing for each benchmark. Results are based on Qwen2.5-7B-Instruct, with a cost ratio of 1:100.}
  \label{fig:rv_100}
% \vspace{-.15in}
\end{figure*}

Figure \ref{fig:rv_25}, \ref{fig:rv_50} and \ref{fig:rv_100} present comparison plots of cost-accuracy(100) between pre-generation routing and cascade routing across multiple benchmarks on Qwen2.5-7B-Instruct, with cost ratios of 1:25, 1:50, and 1:100, respectively.

\subsection{Different Capabilities}
\begin{figure*}[t]
  \centering
  % \vspace{-.15in}
  \includegraphics[width=0.95\textwidth]{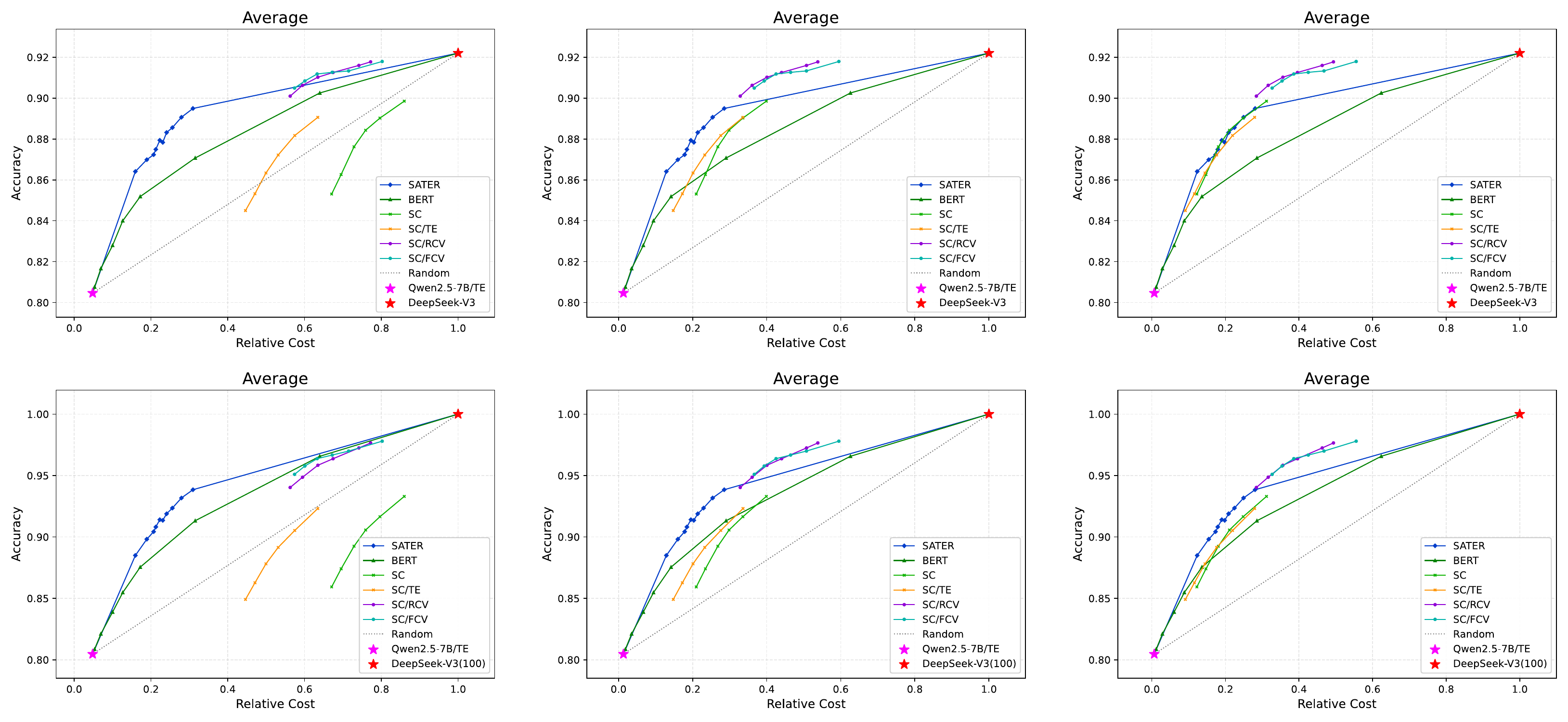}
  % \vspace{-.3in}
  \caption{Comparison between the cost-accuracy plots and the cost-accuracy(100) plots, averaged across all benchmarks. The upper section presents the actual results from the LLM (DeepSeek-V3), while the lower section assumes LLMs deliver optimal responses. The results are based on Qwen2.5-7B-Instruct, with three subplots illustrating cost ratios of 1:13.75, 1:50, and 1:100 (from left to right).}
  \label{fig:dk_13_50_100}
% \vspace{-.15in}
\end{figure*}

In Figure \ref{fig:dk_13_50_100}, we compare the impact of large language models (LLMs) with varying capabilities—specifically, the actual results from DeepSeek-V3 versus the assumption that LLMs always deliver optimal responses—on the performance of prefix routing and cascade routing.

\subsection{Different Temperature Settings}
\begin{figure*}[t]
  \centering
  % \vspace{-.15in}
  \includegraphics[width=0.95\textwidth]{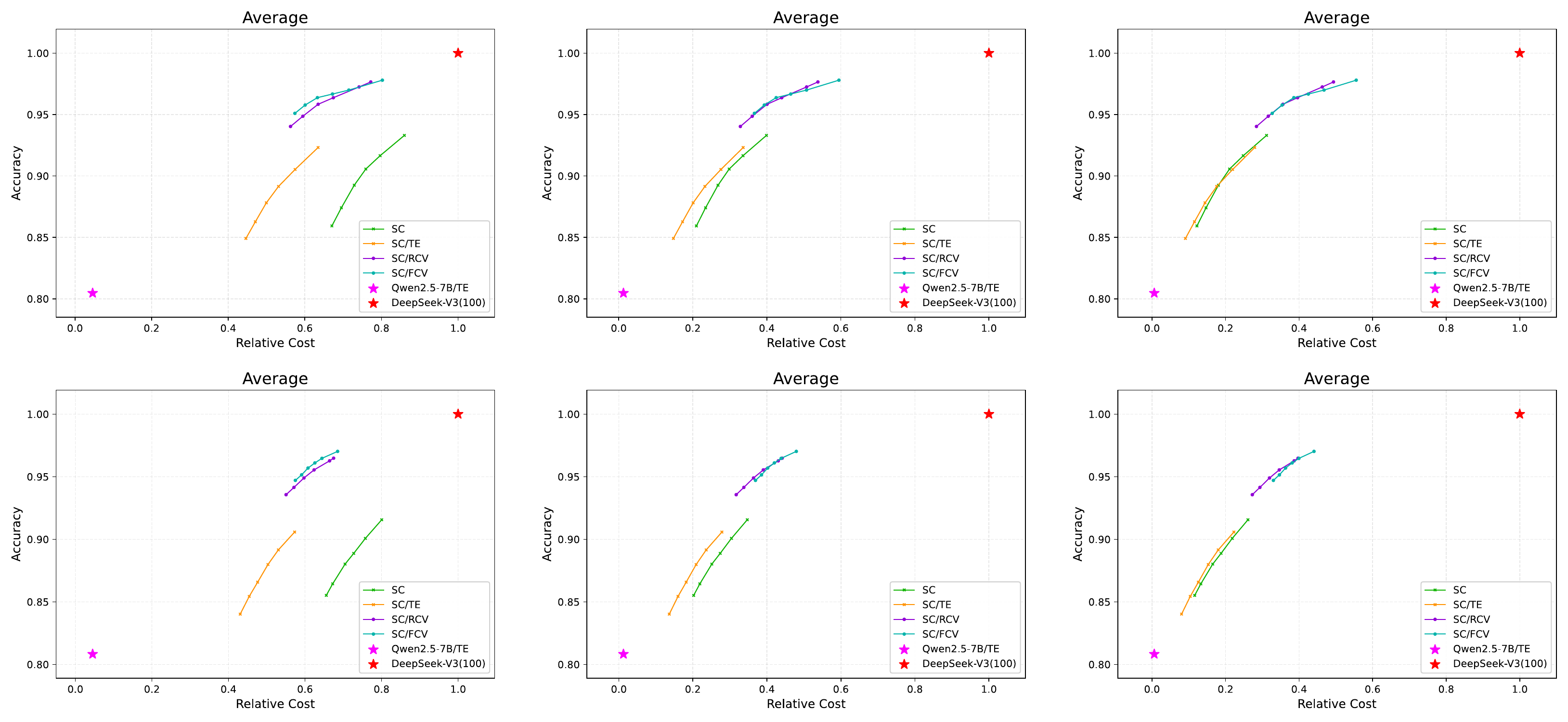}
  % \vspace{-.3in}
  \caption{Comparison between different temperature settings, averaged across all benchmarks. The upper section shows results at temperature 0.7, while the lower section displays results at temperature 0.3. The results are based on Qwen2.5-7B-Instruct, with three subplots illustrating cost ratios of 1:13.75, 1:50, and 1:100 (from left to right).}
  \label{fig:t03}
% \vspace{-.15in}
\end{figure*}

We show the comparison under different temperature settings in Figure \ref{fig:t03}.

\section{Additional Tables}
\subsection{AGL and AROL Results with Different Thresholds}
We present AGL and AROL results with threshold $\tau = 1.0$ in Table \ref{tab: agl-arol-threshold-1}. In mathematical reasoning tasks, a threshold of 0.6 usually yields satisfactory results. For other types of tasks, further performance improvements can be achieved by adjusting the threshold within the range of 0.6 to 1.0.

\subsection{Effectiveness of Long to Short Training}
\label{sec:lts-effectiveness}
Table \ref{tab: efficiency-comparison} presents the efficiency comparison between SATER and TokenSkip on the GSM8K and MATH-500 benchmarks. In Table \ref{tab: accuracy-tokens}, we report the average accuracy and average Chain-of-Thought (CoT) token count across both in-domain and out-of-domain datasets. Table \ref{tab: delta-accuracy-tokens} displays the average accuracy percentage change ($\Delta$\%) and the average CoT token count percentage change ($\Delta$\%) across in-domain and out-of-domain datasets.

\begin{table}[!ht]
\centering
\resizebox{\columnwidth}{!}{ % 自动缩放至单栏宽度
\begin{tabular}{l *{4}{c}}
\toprule
\midrule
\multirow{2}{*}{\textbf{Method}} & \multicolumn{2}{c}{\textbf{GSM8K}} & \multicolumn{2}{c}{\textbf{MATH-500}} \\
\cmidrule(lr){2-3} \cmidrule(lr){4-5}
& \textbf{Accuracy} & \textbf{Tokens} & \textbf{Accuracy} & \textbf{Tokens} \\
\midrule
SATER & 83.0 & 117 & 41.2 & 212 \\
TokenSkip (0.5) & 78.2 & 113 & 40.2 & 292 \\
\midrule
\bottomrule
\end{tabular}
}
\caption{Efficiency comparison between SATER and TokenSkip on GSM8K and MATH-500 benchmarks.}
\label{tab: efficiency-comparison}
\end{table}

\section{Implementation Details}
\label{sec: implementation-details}
All sampling is conducted using the vLLM framework \citep{kwon2023efficient} with a maximum generation length of 1024 tokens and sampling parameters set to temperature = 0.7 and top\_p = 1.0. For DeepSeek-V3, on the GSM8K and MATH-500 datasets, we set temperature = 0 and max\_length = 8192, while other datasets use official default parameters.

The training process utilizes the LLaMA-Factory framework \citep{zheng2024llamafactory} and is performed on four NVIDIA RTX 3090 GPUs. For long-to-short training, we employ the DPO method with LoRA fine-tuning (rank = 8, alpha = 16, dropout = 0.1) for one epoch, using the AdamW optimizer (learning rate = 1e-4) with a cosine learning rate schedule (10\% warmup ratio). The per-device batch size is set to 1 with 4 gradient accumulation steps, a length limit of 1024 tokens, a Sigmoid preference loss ($\beta$ = 1.0), and an auxiliary loss coefficient of 0.2. For refusal training, we use a pure SFT task with the same LoRA fine-tuning parameters as the DPO stage, excluding preference loss parameters, and train for one epoch.

\clearpage
\newpage

\begin{table*}[t]
\centering
\footnotesize
\resizebox{\textwidth}{!}{%
\adjustbox{max width=\textwidth}{
\begin{tabular}{ll *{14}{c}}
\toprule
\midrule
\multirow{2}{*}{\textbf{Model}} & \multirow{2}{*}{\textbf{Method}} & \multicolumn{2}{c}{\textbf{MMLU}} & \multicolumn{2}{c}{\textbf{MATH-500}} & \multicolumn{2}{c}{\textbf{GSM8K}} & \multicolumn{2}{c}{\textbf{ARC\_C}} & \multicolumn{2}{c}{\textbf{ReClor}} & \multicolumn{2}{c}{\textbf{ARC\_E}} & \multicolumn{2}{c}{\textbf{Average}} \\
\cmidrule(lr){3-4} \cmidrule(lr){5-6} \cmidrule(lr){7-8} \cmidrule(lr){9-10} \cmidrule(lr){11-12} \cmidrule(lr){13-14} \cmidrule(lr){15-16}
& & \textbf{AGL} & \textbf{AROL} & \textbf{AGL} & \textbf{AROL} & \textbf{AGL} & \textbf{AROL} & \textbf{AGL} & \textbf{AROL} & \textbf{AGL} & \textbf{AROL} & \textbf{AGL} & \textbf{AROL} & \textbf{AGL} & \textbf{AROL}\\
\midrule
\multirow{4}{*}{\makecell{Llama-3.1-\\8B-Instruct}} & SC & 232 & 138 & 310 & 327 & 192 & 180 & 200 & 110 & 184 & 113 & 186 & 105 & 217 & 162\\
& SC/TE & 87 & 50 & 149 & 125 & 127 & 94 & 75 & 32 & 107 & 59 & 63 & 27 & 101 & 65 \\
& SC/RCV & \underline{64} & \underline{1} & \underline{148} & \underline{8} & \underline{122} & \underline{28} & \underline{60} & \underline{2} & \underline{95} & \underline{3} & \underline{47} & \underline{1} & \underline{89} & \underline{7}\\
& SC/FCV & \textbf{63} & \textbf{1} & \textbf{126} & \textbf{2} & \textbf{117} & \textbf{11} & \textbf{58} & \textbf{1} & \textbf{93} & \textbf{1} & \textbf{46} & \textbf{1} & \textbf{84} & \textbf{3}\\
\midrule
\multirow{4}{*}{\makecell{Qwen2.5-\\7B-Instruct}} & SC & 218 & 243 & 503 & 597 & 324 & 312 & 176 & 159 & 390 & 309 & 146 & 126 & 293 & 291\\
& SC/TE & 129 & 142 & 375 & 396 & 258 & 249 & 96 & 88 & 216 & 176 & 79 & 79 & 192 & 188\\
& SC/RCV & \underline{109} & \underline{5} & \underline{312} & \underline{19} & \underline{234} & \underline{35} & \underline{96} & \underline{3} & \underline{175} & \underline{4} & \underline{80} & \underline{2} & \underline{168} & \underline{11} \\
& SC/FCV & \textbf{106} & \textbf{1} & \textbf{278} & \textbf{3} & \textbf{225} & \textbf{14} & \textbf{94} & \textbf{1} & \textbf{168} & \textbf{5} & \textbf{80} & \textbf{1} & \textbf{159} & \textbf{4} \\
\midrule
\multirow{4}{*}{\makecell{Qwen2.5-\\3B-Instruct}} & SC & 312 & 267 & 504 & 542 & 338 & 306 & 291 & 228 & 455 & 380 & 260 & 216 & 360 & 323\\
& SC/TE & 161 & 151 & 359 & 362 & 231 & 201 & 145 & 105 & 325 & 246 & 113 & 94 & 222 & 193\\
& SC/RCV & \underline{108} & \underline{2} & \underline{259} & \underline{6} & \underline{201} & \underline{16} & \underline{108} & \underline{5} & \underline{272} & \underline{8} & \underline{87} & \underline{4} & \underline{173} & \underline{7} \\
& SC/FCV & \textbf{107} & \textbf{1} & \textbf{256} & \textbf{2} & \textbf{193} & \textbf{4} & \textbf{104} & \textbf{2} & \textbf{278} & \textbf{5} & \textbf{86} & \textbf{2} & \textbf{171} & \textbf{3}\\
\midrule
\bottomrule
\end{tabular}
}
}
\caption{AGL and AROL results with threshold $\tau = 1.0$.}
\label{tab: agl-arol-threshold-1}
\end{table*}

\begin{table*}[t]
\centering
% \footnotesize
\resizebox{\textwidth}{!}{%
\adjustbox{max width=\textwidth}{
\begin{tabular}{ll *{16}{c}}
\toprule
\midrule
\multirow{2}{*}{\textbf{Model}} & \multirow{2}{*}{\textbf{Method}} & \multicolumn{2}{c}{\textbf{MMLU}} & \multicolumn{2}{c}{\textbf{MATH-500}} & \multicolumn{2}{c}{\textbf{GSM8K}} & \multicolumn{2}{c}{\textbf{ARC\_C}} & \multicolumn{2}{c}{\textbf{ReClor}} & \multicolumn{2}{c}{\textbf{ARC\_E}} & \multicolumn{2}{c}{\textbf{Average}} \\
\cmidrule(lr){3-4} \cmidrule(lr){5-6} \cmidrule(lr){7-8} \cmidrule(lr){9-10} \cmidrule(lr){11-12} \cmidrule(lr){13-14} \cmidrule(lr){15-16}
& & \textbf{Acc} & \textbf{Tokens} & \textbf{Acc} & \textbf{Tokens} & \textbf{Acc} & \textbf{Tokens} & \textbf{Acc} & \textbf{Tokens} & \textbf{Acc} & \textbf{Tokens} & \textbf{Acc} & \textbf{Tokens} & \textbf{Acc} & \textbf{Tokens} \\
\midrule
\multirow{2}{*}{\makecell{Llama-3.1-\\8B-Instruct}} & Original & 70.1 & 186 & 47.2 & 457 & 85.1 & 174 & 83.3 & 140 & 62.8 & 130 & 89.8 & 128 & 73.1 & 202.5 \\
& SATER & 68.1 & 67 & 41.2 & 212 & 83.0 & 117 & 81.7 & 44 & 60.6 & 86 & 89.2 & 34 & 70.6 & 93.3 \\
\midrule
\multirow{2}{*}{\makecell{Qwen2.5-\\7B-Instruct}} & Original & 72.5 & 180 & 73.0 & 530 & 92.5 & 280 & 89.1 & 121 & 74.0 & 321 & 92.5 & 95 & 82.3 & 254.5 \\
& SATER & 71.5 & 111 & 69.2 & 392 & 91.0 & 226 & 88.9 & 73 & 73.2 & 163 & 92.8 & 58 & 81.1 & 170.5 \\
\midrule
\multirow{2}{*}{\makecell{Qwen2.5-\\3B-Instruct}} & Original & 69.1 & 262 & 63.0 & 555 & 86.0 & 297 & 83.9 & 214 & 62.6 & 415 & 90.8 & 176 & 75.9 & 319.8 \\
& SATER & 68.6 & 138 & 61.8 & 391 & 81.5 & 212 & 84.4 & 94 & 63.2 & 266 & 90.7 & 69 & 75.0 & 195.0 \\
\midrule
\bottomrule
\end{tabular}
}
}
\caption{Average accuracy and average CoT token count (Tokens) across both in-domain and out-of-domain datasets.}
\label{tab: accuracy-tokens}
\end{table*}

\begin{table*}[!ht]
\centering
% \footnotesize
\resizebox{\textwidth}{!}{%
\adjustbox{max width=\textwidth}{
\begin{tabular}{ll *{14}{c}}
\toprule
\midrule
\multirow{2}{*}{\textbf{Model}} & \multirow{2}{*}{\textbf{Method}} & \multicolumn{2}{c}{\textbf{MMLU}} & \multicolumn{2}{c}{\textbf{MATH-500}} & \multicolumn{2}{c}{\textbf{GSM8K}} & \multicolumn{2}{c}{\textbf{ARC\_C}} & \multicolumn{2}{c}{\textbf{ReClor}} & \multicolumn{2}{c}{\textbf{ARC\_E}} & \multicolumn{2}{c}{\textbf{Average}} \\
\cmidrule(lr){3-4} \cmidrule(lr){5-6} \cmidrule(lr){7-8} \cmidrule(lr){9-10} \cmidrule(lr){11-12} \cmidrule(lr){13-14} \cmidrule(lr){15-16}
& & \textbf{$\Delta$Acc} & \textbf{$\Delta$Tokens} & \textbf{$\Delta$Acc} & \textbf{$\Delta$Tokens} & \textbf{$\Delta$Acc} & \textbf{$\Delta$Tokens} & \textbf{$\Delta$Acc} & \textbf{$\Delta$Tokens} & \textbf{$\Delta$Acc} & \textbf{$\Delta$Tokens} & \textbf{$\Delta$Acc} & \textbf{$\Delta$Tokens} & \textbf{$\Delta$Acc} & \textbf{$\Delta$Tokens} \\
\midrule
Llama-3.1-8B & SATER & -2.0 & -64.2 & -6.0 & -53.6 & -2.1 & -32.5 & -1.6 & -68.2 & -2.2 & -34.4 & -0.6 & -73.2 & -2.4 & -54.4 \\
\midrule
Qwen2.5-7B & SATER & -1.0 & -38.3 & -3.8 & -26.1 & -1.5 & -19.0 & -0.2 & -39.6 & -0.8 & -49.0 & +0.3 & -38.3 & -1.2 & -35.1 \\
\midrule
Qwen2.5-3B & SATER & -0.5 & -47.3 & -1.2 & -29.4 & -4.5 & -28.6 & +0.5 & -56.1 & +0.6 & -36.1 & -0.1 & -60.6 & -0.9 & -43.0 \\
\midrule
\bottomrule
\end{tabular}
}
}
\caption{Average accuracy percentage change ($\Delta$\%) and average CoT token count percentage change ($\Delta$\%) across both in-domain and out-of-domain datasets, where a negative value indicates a decrease and a positive value indicates an increase.}
\label{tab: delta-accuracy-tokens}
\end{table*}

\begin{table*}[!ht]
\centering
\resizebox{\textwidth}{!}{%
\begin{tabular}{ll *{12}{c}}
\toprule
\midrule
\multirow{2}{*}{\textbf{Model}} & \multirow{2}{*}{\textbf{Method}} & \multicolumn{2}{c}{\textbf{MMLU}} & \multicolumn{2}{c}{\textbf{MATH-500}} & \multicolumn{2}{c}{\textbf{GSM8K}} & \multicolumn{2}{c}{\textbf{ARC\_C}} & \multicolumn{2}{c}{\textbf{ReClor}} & \multicolumn{2}{c}{\textbf{ARC\_E}} \\
\cmidrule(lr){3-4} \cmidrule(lr){5-6} \cmidrule(lr){7-8} \cmidrule(lr){9-10} \cmidrule(lr){11-12} \cmidrule(lr){13-14}
& & \textbf{ToA-100} & \textbf{ToGR} & \textbf{ToA-100} & \textbf{ToGR} & \textbf{ToA-100} & \textbf{ToGR} & \textbf{ToA-100} & \textbf{ToGR} & \textbf{ToA-100} & \textbf{ToGR} & \textbf{ToA-100} & \textbf{ToGR} \\
\midrule
\multirow{5}{*}{\makecell{Qwen2.5-7B}} 
& SATER & \textbf{0.692} & \textbf{0.533} & \textbf{0.770} & \textbf{0.810} & \textbf{0.708} & \textbf{0.461} & \textbf{0.631} & \textbf{0.294} & \textbf{0.639} & \textbf{0.385} & \textbf{0.593} & \textbf{0.201} \\
& FrugalGPT & 0.654 & 0.428 & 0.712 & 0.636 & 0.657 & 0.348 & 0.553 & 0.120 & 0.553 & 0.146 & 0.540 & 0.087 \\
& BERT & 0.629 & 0.360 & 0.690 & 0.568 & 0.532 & 0.070 & 0.549 & 0.111 & 0.554 & 0.150 & 0.529 & 0.062 \\
& Automix & 0.499 & $-$0.002 & 0.582 & 0.246 & 0.567 & 0.148 & 0.490 & $-$0.021 & 0.472 & $-$0.076 & 0.458 & $-$0.090 \\
& Margin Sampling & 0.531 & 0.085 & 0.505 & 0.014 & 0.511 & 0.024 & 0.567 & 0.151 & 0.535 & 0.098 & 0.556 & 0.121 \\
\midrule
\bottomrule
\end{tabular}
}
\caption{Additional ToA-100 and ToGR results across in-domain and out-of-domain datasets. Bold indicates the best.}
\label{tab:toa_togr}
\end{table*}

% \clearpage
% \onecolumn
\end{document}